\documentclass[twocolumn,showpacs,preprintnumbers,amsmath,amssymb]{revtex4}
\usepackage{graphics}
\usepackage{epsfig}
\begin{document}

\title{Controlled Hopf Bifurcation of a storage-ring free-electron laser}
\author{Giovanni De Ninno$^{1}$\thanks{giovanni.deninno@elettra.trieste.it},
Duccio Fanelli$^{2}$\thanks{fanelli@et3.cmb.ki.se}}
\affiliation{1. Sincrotrone Trieste, 34012 Trieste, Italy \\
2. Cell and Molecular Biology Department, Karolinska Institute,
SE-171 77 Stockholm, Sweden}

\date{\today}

\begin{abstract}
Local bifurcation control is a topic of fundamental importance in
the field of nonlinear dynamical systems. We discuss an original
example within the context of storage-ring free-electron laser
physics by presenting a new model that enables analytical insight
into the system dynamics. The transition between the stable and
the unstable regimes, depending on the temporal overlapping
between the light stored in the optical cavity and the electrons
circulating into the ring, is found to be a Hopf bifurcation. A
feedback procedure is implemented and shown to provide an
effective stabilization of the unstable steady state.
\end{abstract}



\maketitle

Transition from stability to chaos is a common characteristic of
many physical and biological systems \cite{phys,bio}. Within this
context, local bifurcation control is a topic of paramount
importance especially for those systems in which stability is a
crucial issue. This is, for example, the case for conventional and
non conventional light sources, such as Storage-Ring Free-Electron
Lasers (SRFELs), commonly employed in various scientific
applications \cite{appl2}. In a SRFEL \cite{colson}, the physical
mechanism responsible for light emission and amplification is the
interaction between a relativistic electron beam and the
magnetostatic periodic field of an undulator. Due to the effect of
the magnetic field the electrons emit synchrotron radiation, known
as spontaneous emission. The light produced by the electron beam
is stored in an optical cavity and amplified during successive
turns of the particles in the ring. A given temporal detuning,
i.e. a difference between the electron beam revolution period and
the round trip of the photons inside the optical cavity, leads to
a cumulative delay between the electrons and the laser pulses: the
laser intensity may then appear as ``continuous wave (cw)'' (for a
weak or strong detuning) or show a stable pulsed behavior (for an
intermediate detuning amount) \cite{detusaco, detuvsor}. The
achievement of a large and stable ``cw'' zone is a crucial issue,
of fundamental importance for experimental applications. In this
Letter, we characterize the transition between stable the unstable
regimes as a Hopf bifurcation. This result allows one to establish
a formal bridge with the field of conventional lasers and to adopt
the universal techniques of control theory to enlarge the stable
signal region. We develop this idea by introducing a new model
which reveals to be particularly suitable for analytic
investigations.
\\
The longitudinal dynamics of a SRFEL can be described by a system
of rate equations accounting for the coupled evolution of the
electromagnetic field and of the longitudinal parameters of the
electron bunch \cite{bill1,gdmec}. The temporal profile of the
laser intensity, $y_n$, is updated at each pass, $n$, inside the
optical cavity according to:
\begin{equation}
\label{pass-pass} y_{n+1} (\tau) = R^2 y_n (\tau-\epsilon) \left[
1 + g_{n}(\tau) \right] +i_s(\tau),
\end{equation}
where $\tau$ is the temporal position of the electron bunch
distribution with respect to the centroid; $R$ is the mirror
reflectivity; the detuning parameter $\epsilon$ is the difference
between the electrons revolution period (divided by the number of
bunches) and the period of the photons inside the cavity;
$i_s(\tau)$ accounts for the profile of the spontaneous emission
of the optical klystron \cite{kly}. Assuming that the saturation
is achieved when the peak gain is equal to the cavity losses, $P$,
the FEL gain $g_{n}(\tau)$ is given by \cite{bill1, gdmec}:
\begin{equation}
\label{gain01} g_{n}(\tau) =g_i \frac{\sigma_0}{\sigma_{n}}
\left[\frac{P \sigma_e }{g_i \sigma_0 }\right]^
{\frac{\sigma_{n}^2 - \sigma_0^2}{\gamma}} \exp \left[
-\frac{\tau^2}{2 \sigma_{\tau,n}^2} \right]
\end{equation}
where $g_i$ and $\sigma_{0}$ are the initial (laser-off) peak gain
and beam energy spread, $\sigma_{n}$ and $\sigma_{{\tau},n}$ are
the energy spread and the bunch length after the n$th$
light-electron beam interaction, and $\gamma = \sigma_{e}^2 -
\sigma_{0}^2 $. Note that equation (\ref{gain01}) refers to the
case of  SRFELs implemented on an optical klystron. The evolution
of the laser-induced electron-beam energy spread is governed by
the following equation:
\begin{equation}
\label{Sigma}
 \sigma_{n+1}^2=\sigma_{n}^2 + \frac{2 \Delta
T}{\tau_{s}}(\gamma I_{n} + \sigma_{0}^2 - \sigma_{n}^2).
\end{equation}
Here $\sigma_e$ is the equilibrium value (i.e. that reached at the
laser saturation) of the energy spread at the perfect tuning and
$\Delta T$ is the bouncing period of the laser inside the optical
cavity; $I_{n}=\int^{\infty}_{-\infty} y_{n}(\tau) d\tau$ is the
laser intensity normalized to its equilibrium value (i.e. the
saturation value for $\epsilon =0$) and $\tau_s$ stands for the
characteristic time of the damped oscillation of electrons in
their longitudinal phase-space. Equations (\ref{pass-pass}),
(\ref{gain01}) and (\ref{Sigma}) are shown to reproduce
quantitatively the experimental results \cite{gdmec}. In
particular, the laser intensity displays a stable ``cw'' behavior
for small amount of detuning, while a pulsed regime is found for
$\epsilon$ larger than a certain critical threshold, $\epsilon_c$.
This model represents the starting point of our analysis.

\begin{figure}[h!]
\resizebox{0.40\textwidth}{!}{\includegraphics{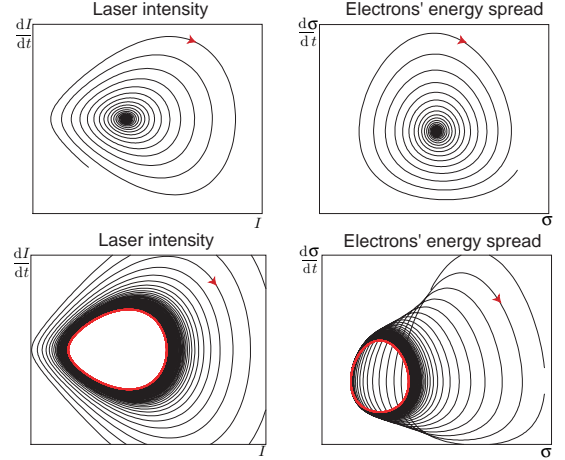}} \vskip
.5truecm \caption{\label{bifPRL} \em Phase-space portraits. Left
column: $\dot{I}$ is plotted versus $I$. Right column:
$\dot{\sigma}$ versus $\sigma$. The top panels refer to $\epsilon
= 0.1fs < \epsilon_c$, the bottom ones to $\epsilon=1.3fs >
\epsilon_c$. Simulations have been performed using the case of the
Super-ACO FEL as reference. The values of the relevant parameters
are: $\Delta T=120$ ns, $\tau_s= 8.5$ ms,
$\sigma_0=5\cdot10^{-4}$, $\sigma_e/\sigma_0 = 1.5$, $\Omega=14$
kHz, $g_i=2\%$, $P=0.8\%$, $I_s=1.4\cdot 10^{-8}$. }
\end{figure}

Equation (\ref{pass-pass}) characterizes the evolution of the
statistical parameters of the laser distribution: by assuming a
specific form for the profile, it is in principle possible to make
explicit the evolution of each quantity. For this purpose, we put
forward the assumption of a Gaussian laser profile and compute the
first three moments. The details of the calculations are given
elsewhere \cite{elettra}. In addition, it is shown that for
$\epsilon$ spanning the central ``cw'' zone, the quantities
$(\sigma_{l,n}/\sigma_{\tau,n})^2$ and $[(\tau_n
+\epsilon)/\sigma_{\tau,n}]^2$ are small. Hence, a Taylor series
expansion is performed and second order terms neglected. Finally,
by approximating finite differences with differentials, the
following continuous system is found:

\begin{equation}
\label{sistemadiff} \left\{
\begin{array}{l}
\displaystyle{\frac{\mbox{d}\sigma}{\mbox{d}t}=\frac{\alpha_1}{\Delta
T} \frac{1}{2 \sigma} \left[\alpha_2I+1-\sigma^2\right]}
\\
\\
\displaystyle{\frac{\mbox{d}I}{\mbox{d}t}=\frac{R^2 I}{\Delta
T}\left[-\frac{P}{R^2}+\frac{g_i \alpha_3}{2
\sigma^3}\alpha_4^{\frac{\sigma^2-1}{\alpha_2}}
\left(\frac{2\sigma^2}{\alpha_3}-\sigma_{l}^2-
\hat{\tau}^2\right)\right]+\frac{I_s}{\Delta T}}
\\
\\
\displaystyle{\frac{\mbox{d}\tau}{\mbox{d}t}=-\frac{\tau}{\Delta T
}+\frac{\hat{\tau}}{\Delta T
}\left[1-\frac{g_i}{\sigma}\alpha_3\alpha_4^{\frac{\sigma^2-1}{\alpha_2}}
\frac{\sigma_{l}^2}{\sigma^2}\right]}
\\
\\
\displaystyle{\frac{\mbox{d}\sigma_{l}}{\mbox{d}t}=-\frac{1}{\Delta
T}\frac{g_i}{2}\alpha_3\alpha_4^{\frac{\sigma^2-1}{\alpha_2}}
\frac{\sigma_{l}^3}{\sigma^3} +\frac{1}{\Delta T }\frac{I_s}{I}
\frac{1}{2
\sigma_{l}}\left(\frac{\sigma^2}{\alpha_3}+\tau^2\right)},
\end{array}
\right.
\end{equation}

where $\hat{\tau}=\tau+\epsilon$ and

\begin{equation}
\label{alpha1} \alpha_1=\frac{2\Delta T}{\tau_s},~~~~~
\alpha_2=\frac{\sigma_e^2-\sigma_0^2}{\sigma_0^2},
\end{equation}
\begin{equation}
\label{alpha3}
\alpha_3=\left(\frac{\Omega}{\sigma_0\alpha}\right)^2,~~~
\alpha_4=\frac{P\sigma_e}{g_i\sigma_0}~.
\end{equation}
Here $\Omega$ represents the oscillation frequency of the
electrons in their longitudinal phase-space and $\alpha$, the
momentum compaction factor, is a characteristic parameter of the
storage ring. Note the redefinition of $\sigma$ which is from
hereon normalized to $\sigma_0$. Although in approximate form,
system (\ref{sistemadiff}) still captures the main features of the
longitudinal SRFEL dynamics. In particular, the transition from
the ``cw'' regime to the unstable (pulsed) steady state occurs for
a temporal detuning which is close to the one found in the
framework of the exact formulation, hence to the experimental
value. However, system (\ref{sistemadiff}) fails in reproducing
the correct behavior when the transition to the lateral ``cw''
zone is approached.  In Figure \ref{bifPRL} phase-space portraits
for both the laser intensity and the beam energy spread are
plotted for different values of $\epsilon$. Limit cycles are
observed when $\epsilon>\epsilon_c$. For smaller values of
$\epsilon$, the variables converge asymptotically to a stable
fixed point. The latter can be analytically characterized, thus
allowing one to relate the electron-beam energy spread, intensity,
centroid position and rms value of the laser distribution, to the
light-electron beam detuning. Through a stability analysis it is
also possible to determine the threshold value $\epsilon_c$. To
our knowledge this study represents the first attempt to fully
characterize the detuned SRFEL dynamics.

\vskip .5truecm

\begin{figure}[t]
\resizebox{0.40\textwidth}{!}{\includegraphics{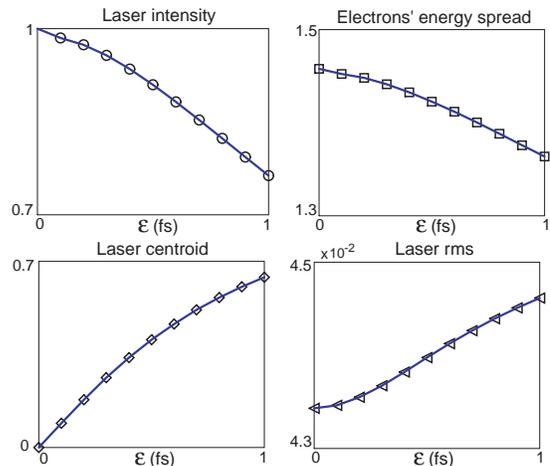}} \vskip
-.5truecm \caption{\label{punti} \em The fixed points are plotted
as function of the detuning parameter $\epsilon$. Top left panel:
Normalized laser intensity. Top right panel: Normalized
electron-beam energy spread. Bottom left panel: Laser centroid.
Bottom right panel: rms value of the laser distribution. Symbols
refer to the simulations, while the solid line stands for the
analytic calculation. The list of parameters is enclosed in the
caption of Figure \ref{bifPRL}.}
\end{figure}

The fixed points
$\left(\overline{I},~\overline{\sigma},~\overline{\tau},~\overline{\sigma_l}\right)$
are found by imposing $\frac{dI}{dt} = \frac{d\sigma}{dt} =
\frac{d\tau}{dt} =  \frac{d\sigma_l}{dt} = 0$~ in
(\ref{sistemadiff}), and solving the corresponding system. Assume
hereon $\epsilon>0$, being the scenario for $\epsilon<0$
completely  equivalent. After some algebraic calculations, the
following relations are found:
\begin{equation}
\label{ifix} \overline{I}=\frac{\overline{\sigma}^2-1}{\alpha_2},
\end{equation}
\begin{equation}
\label{taufix}
\overline{\tau}=\left\{\frac{1}{2}\left[-\frac{\overline{\sigma}^2}{\alpha_3}+
\sqrt{\left(\frac{\overline{\sigma}^2}{\alpha_3}\right)^2+4\epsilon^2A}\right]\right\}^{\frac{1}{2}},
\end{equation}
\begin{equation}
\label{sigmalas}
\overline{\sigma_l}=\left\{\frac{I_s}{2g_i\alpha_3}\alpha_4^{\frac{1-\overline{\sigma}^2}{\alpha_2}}
\alpha_2\frac{\overline{\sigma}^3}{\overline{\sigma}^2-1}\left[\frac{\overline{\sigma}^2}{\alpha_3}
+\sqrt{\left(\frac{\overline{\sigma}^2}{\alpha_3}\right)^2+4\epsilon^2A}\right]\right\}^{\frac{1}{4}},
\end{equation}
where
\begin{equation}
\label{sigmalequi2}
A=\frac{\overline{\sigma}^3\left(\overline{\sigma}^2-1\right)}{\alpha_2I_s}
\frac{\alpha_4^{\frac{1-\overline{\sigma}^2}{\alpha_2}}}{g_i\alpha_3}.
\end{equation}
These relations link the equilibrium values of $\overline{I},
\overline{\tau}, \overline{\sigma_l}$ to $\overline{\sigma}$. The
quantity $\overline{\sigma}$ is found from the following implicit
equation:

\begin{equation}
\label{implicit_sigma} \frac{g_i}{{\overline{\sigma}}}
\alpha_4^{\frac{\overline{\sigma}^2-1}{\alpha_2}} \left[ 1 -
\frac{1}{2} \frac{\alpha_3}{\overline{\sigma}^2}
\left(\overline{\sigma_l}^2 + (\overline{\tau}+\epsilon)^2 \right)
\right]= \frac{P}{R^2},
\end{equation}
where $\overline{\sigma_l}$ and $\overline{\tau}$ are respectively
given by (\ref{sigmalas}) and  (\ref{taufix}). For any given value
of the detuning $\epsilon$, equation (\ref{implicit_sigma}) can be
solved numerically, by using a standard bisection method. The
estimates of $\overline{\sigma}$ are then inserted in equations
(\ref{ifix}), (\ref{taufix}), (\ref{sigmalas}), to compute the
corresponding values of
$\overline{I},~\overline{\tau},~\overline{\sigma_l}$. Results of
the calculations (solid line) and direct numerical simulations
using the system (\ref{sistemadiff}) (symbols) are compared in
Figure \ref{punti}, displaying remarkably good agreement. It is
worth stressing that, by means of a perturbative analysis, a
closed analytical expression for $\overline{\sigma}$ as a function
of $\epsilon$ is also found. The details of the quite cumbersome
calculations are given elsewhere \cite{elettra}.

As a validation of the preceding analysis, we consider the case of
perfect tuning, i.e. $\epsilon=0$, and compare our estimate for
the laser induced energy  spread $\overline{\sigma_l}$ to the
value $(\overline{\sigma_l})_{sm}$, derived in the context of the
widely used super-modes approach \cite{sm1}. Both theoretical
predictions are then compared to experiments performed on the
SuperACO and Elettra FELs. Results are given in Table \ref{tab1}:
the improvement of the calculation based on equation
(\ref{sigmalas}) is clearly shown. The results of Table \ref{tab1}
indicate that both $\overline{\sigma_l}$ and
$(\overline{\sigma_l})_{sm}$ are smaller than the experimental
values. This is probably due to the fact that the models neglect
the effect of the microwave instability \cite{mic} resulting from
the electron beam interaction with the ring environment (e.g. the
metallic wall of the vacuum chamber). In the case of Elettra the
situation is complicated by the presence of a ``kick-like''
instability (having a characteristic frequency of 50 Hz) which
periodically switches off the laser preventing the attainment of a
stable ``cw'' regime \cite{ele}.

\begin{table}[h!]
\begin{center}
\vspace{5mm}
\begin{tabular}{ |c| |c| |c| }
\hline
&Super ACO&Elettra\\
\hline\hline
$\overline{\sigma_l}$ (ps) &5& 2\\
\hline
($\overline{\sigma_l})_{sm}$ (ps)&3&1\\
\hline
Experimental values (ps)&$10\pm2$&$5\pm2$\\
\hline
\end{tabular}
\end{center}
\caption{\label{tab1} \em Theoretical widths of the laser pulse
compared to experimental values for the case of the Super-ACO and
Elettra FELs. The experimental setting for the case of Super ACO
(operated at a beam energy of $800$ MeV and at a laser wavelength
of $350$ nm) is that specified in the caption of Figure
\ref{bifPRL}. The analogous parameters for ELETTRA (operated at a
beam energy of $900$ MeV and at a laser wavelength of $250$ nm)
are the following: $\Delta T=216$ ns, $\tau_s= 87$ ms,
$\sigma_0=1\cdot10^{-3}$, $\sigma_e/\sigma_0 = 1.5$, $\Omega=16$
kHz, $g_i=15\%$, $P=7\%$, $I_s=4.3\cdot 10^{-7}$.}
\end{table}

\begin{figure}[t]
\resizebox{0.3\textwidth}{!}{\includegraphics{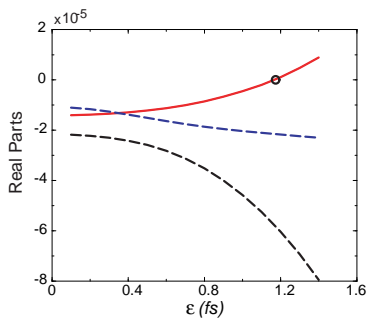}} \vskip
-.5truecm \caption{\label{eig} \em Real part of the eigenvalues of
the Jacobian matrix associated to the system (\ref{sistemadiff})
as a function of the detuning parameter $\epsilon$. The solid line
refers to the complex conjugate eigenvalues. The circle represents
the transition from the stable to the pulsed regime, i.e. the Hopf
bifurcation.}
\end{figure}

The stability of the fixed point $\left[\overline{I}(\epsilon),
\overline{\sigma}(\epsilon), \overline{\tau}(\epsilon),
\overline{\sigma_l}(\epsilon)\right]$ can be determined by
studying the eigenvalues of the Jacobian matrix associated with
the system (\ref{sistemadiff}). The real part of the eigenvalues
as a function of $\epsilon$ is shown in Figure \ref{eig}. The
system is by definition stable when all the real parts are
negative. The transition to an unstable regime occurs when at
least one of them becomes positive. In general, the loss of
stability takes place according to different modalities. Consider
the case of a Jacobian matrix with a pair of complex conjugate
eigenvalues and assume the real parts of all the eigenvalues to be
negative. A Hopf bifurcation occurs when the real part of the two
complex eigenvalues becomes positive, provided the other keep
their signs unchanged \cite{berg}. This situation is clearly
displayed in Figure \ref{eig}, thus allowing to conclude that the
transition between the ``cw'' and the pulsed regime in a SRFEL is
a Hopf bifurcation. The critical detuning, $\epsilon_c$, can be
calculated (open circle in Figure (\ref{eig})) and displays good
agreement with both the simulated data and the experimental value.
A closed relation for $\epsilon_c$ is also found \cite{elettra},
by making use of the analytic expressions for the fixed points.

\begin{figure}[t]
\resizebox{0.40\textwidth}{!}{\includegraphics{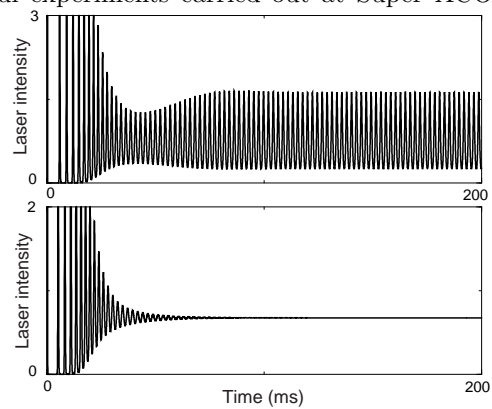}} \vskip
-.5truecm \caption{\label{stabile} \em Behavior of the FEL
(normalized) intensity in absence (upper panel) and in presence
(lower panel) of the derivative control system. The simulations
refer to the case of Super ACO (see caption of Figure \ref{bifPRL}
for the list of the parameters). Here $\epsilon_0=1.3$ fs
$>\epsilon_c$. The stabilization has been achieved using
$\beta=6\cdot10^{-3}$. Here, $\beta_c \simeq 5\cdot10^{-4}$.}
\end{figure}

Having characterized the transition from the stable to the
unstable steady state in term of Hopf bifurcation opens up
interesting perspectives to stabilize the signal and dramatically
improve the system performance. In order to maintain the
laser-electron beam synchronism and avoid the migration towards
one of the unstable pulsed zones of the detuning curve, existing
second-generation SRFELs, such like Super-ACO and UVSOR
\cite{feed1,feed2}, have implemented dedicated control systems.
The idea is to re-adjust periodically the radio-frequency, thus
dynamically confining the laser in the central ``cw'' zone. Even
though generally suitable for second-generation SRFELs, these
systems are inappropriate for more recent devices, such as ELETTRA
and DUKE. The latters are indeed characterized by a much narrower
region of stable signal only occasionally experimentally observed
\cite{ele}, making a priori impossible to pursuit the former
strategy. On the contrary, the approach here discussed exploits an
{\it universal} property of SRFELs, thus allowing to overcome the
limitations of other schemes. The procedure consists in
introducing a specific self-controlled (closed loop) feedback to
suppress locally the Hopf bifurcation and {\it enlarge} the zone
of stable signal. This is achieved by replacing the constant
detuning with the time-dependent quantity \cite{glorieux}:
\begin{equation}
\label{epsvar} \epsilon (t)=\epsilon_0+\beta \Delta T \dot{I}~,
\end{equation}
which is added to system (\ref{sistemadiff}). Here $\epsilon_0$ is
assumed to be larger that $\epsilon_{c}$: when the control is
switched off, i.e. $\beta=0$, the laser is unstable and displays
periodic oscillations. For $\beta$ larger than a certain
threshold,  $\beta_c$, the oscillations are damped and the laser
behaves as if it were in the ``cw'' region. Note that, as soon as
saturation is reached, $\dot{I}=0$ and, thus, the stable regime is
maintained asymptotically for $\epsilon=\epsilon_0>\epsilon_c$,
i.e. well inside the former unstable zone. The results of the
simulations are represented in Figure \ref{stabile}.

This new theoretical insight sets the ground for experimental
tests \cite{elettra}. In this respect, a significant and
reproducible extension of the stable ``cw'' region using this
technique has been recently achieved at Super ACO \cite{ME}. This
result fully confirms our theoretical predictions.

In conclusion, in this Letter we propose a new approximate model
of a SRFEL. This formulation enables a deep analytical insight
into the system dynamics, allowing one to derive the explicit
dependence of the main laser parameters on the temporal detuning.
Results are fully confirmed by numerical simulations and show
satisfactory agreement with available experimental data. Further,
the transition between the stable and unstable regimes is found to
be a Hopf bifurcation, and the critical detuning $\epsilon_c$  is
calculated explicitly. Finally, we introduced in the model a
derivative feedback that is shown to stabilize the laser intensity
well beyond the threshold $\epsilon_c$. Successful experiments
carried out at Super ACO confirmed our predictions. Preliminary
experiments carried out at ELETTRA have also given encouraging
results.

\end{document}